\documentclass[journal]{IEEEtran}
\usepackage{fancyhdr}
\usepackage{amsmath}
\usepackage{amssymb}
\usepackage{graphicx}
\usepackage{booktabs}
\usepackage{amsfonts}
\usepackage{multirow}
\usepackage{algorithm}
\usepackage{algorithmic}
\usepackage{flushend}
\usepackage{mathrsfs}
\usepackage{color}

\usepackage{exscale}
\usepackage{relsize}

\usepackage{cite}

\newtheorem{theorem}{\textbf{Theorem}}

\ifCLASSINFOpdf
\else
\fi

\begin{document}

\bibliographystyle{IEEEtran} % use IEEEtran.bst style

% paper title
%\title{Spectral Efficiency Analysis for Single-Cell Massive Single-Carrier Spatial Modulation MIMO}
\title{Spectral Efficiency Analysis for Spatial Modulation in Massive MIMO Uplink over Dispersive Channels}

%\author{Author~1,
%        Author~2,
%        Author~3,
%        Author~4
%        and Author~5}

\author{Yue~Sun,~\IEEEmembership{Student~Member,~IEEE},
        Jintao~Wang,~\IEEEmembership{Senior~Member,~IEEE},
        Longzhuang~He,~\IEEEmembership{Student~Member,~IEEE},
        and Jian~Song,~\IEEEmembership{Fellow,~IEEE}
\thanks{
Yue Sun, Jintao Wang, Longzhuang He and Jian Song are with the Department of Electronic Engineering and Tsinghua National Laboratory for Information Science and Technology (TNList), Tsinghua University, Beijing, 100084, China (email: suny15@tsinghua.edu.cn).

This work was supported by the National Natural Science Foundation of China (Grant No. 61471221 and No. 61471219).
}}

% make the title area
\maketitle

\begin{abstract}
Flat-fading channel models are usually invoked for analyzing the performance of massive spatial modulation multiple-input multiple-output (SM-MIMO) systems. However, in the context of broadband SM transmission, the severe inter-symbol-interference (ISI) caused by the frequency-selective fading channels can not be ignored, which leads to very detrimental effects on the achievable system performance, especially for single-carrier SM (SC-SM) transmission schemes. To the best of the author's knowledge, none of the previous researchers have been able to provide a thorough analysis on the achievable spectral efficiency (SE) of the massive SC-SM MIMO uplink transmission. In this context, the uplink SE of single-cell massive SC-SM MIMO system is analyzed, and a tight closed-form lower bound is proposed to quantify the SE when the base station (BS) uses maximum ratio (MR) combining for multi-user detection. The impacts of imperfect channel estimation and transmit antenna (TA) correlations are all considered. Monte Carlo simulations are performed to verify the tightness of our proposed SE lower bound. Both the theoretical analysis and simulation results show that the SE of uplink single-cell massive SC-SM MIMO system has the potential to outperform the uplink SE achieved by single-antenna UEs.
\end{abstract}

\begin{keywords}
massive MIMO; spatial modulation (SM); single-carrier SM (SC-SM); frequency-selective fading; spectral efficiency (SE).
\end{keywords}

\IEEEpeerreviewmaketitle

\section{Introduction}
Recently, with a large number of antennas which are equipped in base station (BS) simultaneously serving different user equipments (UEs) in each cell, massive multiple-input multiple-output (MIMO) system is proposed to achieve a much higher spectral efficiency (SE) comparing with traditional MIMO system, which is considered as a promising technique in next-generation cellular communication networks \cite{Massive_CM,Massive_SPM,Massive_SINR,Massive_Ding}. In addition, with the single radio frequency (RF) chain property, spatial modulation (SM) system is proposed as a novel architecture of MIMO to achieve a higher energy efficiency (EE) than traditional MIMO systems \cite{SM}\cite{SM_mgz}. To simultaneously obtain the benefits of massive MIMO and SM, massive SM-MIMO system is proposed to achieve high SE and EE, in which a huge number of receive antennas (RAs) in base station (BS) and some UEs with more than one transit antennas (TAs) but single RF chain are equipped \cite{Massive_SM_ITA,Massive_SM_TWC,Massive_SM_HLZ}.

However, in traditional SM system, the channel is assumed to be a Rayleigh flat fading channel \cite{SM}\cite{SM_mgz}, so traditional SM system can only be implemented in narrowband scenarios  \cite{SC-SM_Hanzo}. With frequency-selective fading property in dispersive channel, traditional SM system can not be deployed in broadband scenarios, so single-carrier SM (SC-SM) system is proposed to preserve the high EE property of SM system \cite{SC-SM_Hanzo,SC-SM_PIMRC,SC-SM_ZP}. In SC-SM system, symbols are transmitted in frame, and cyclic prefix (CP) \cite{SC-SM_PIMRC} or zero padding (ZP) \cite{SC-SM_ZP} is utilized to eliminate the impact of inter-symbol interference(ISI).

Recently, as the combination of SC-SM and massive MIMO, massive SC-SM MIMO system is proposed to achieve both high SE and EE in broadband scenarios, in which a BS with massive RAs provides service for several UEs utilizing SC-SM scheme for uplink transmission \cite{SC-SM_LS_Mag}\cite{SC-GSM_LS}. However, until now only the multi-user detection of massive SC-SM MIMO system is under studied \cite{SC-GSM_LS}\cite{SC-SM_LS_DT}, and there are no research about the SE analysis of massive SC-SM MIMO system.

Therefore, in this paper, we analyze the achievable SE of uplink single-cell massive SC-SM MIMO system, and a theoretical framework is proposed for deriving the lower bound of SE with different linear combining algorithms in BS. In addition, a novel tight closed-form lower bound of SE in uplink single-cell massive SC-SM MIMO system with maximum ratio (MR) combining is proposed, in which the imperfect channel state information (CSI) is acquired via time-division (TD) orthogonal pilot and zero forcing (ZF) channel estimation algorithm. In this framework, the SE is determined by the mutual information, and the signal-to-interference-plus-noise-ratio (SINR) is derived to calculate the mutual information. With certain algorithms of channel estimation and combining in BS, the closed-form SINR can be derived, then the SE of system can be derived too.

The remainder of this paper is organized as follows. In section \uppercase\expandafter{\romannumeral2}, the system model of uplink single-cell massive SC-SM MIMO, and channel estimation algorithm are introduced. In section \uppercase\expandafter{\romannumeral3}, the framework for SE analysis of uplink single-cell massive SC-SM MIMO system is proposed. In section \uppercase\expandafter{\romannumeral4}, the tight lower bound of SE in uplink single-cell massive SC-SM MIMO system with MR combining is derived. In section \uppercase\expandafter{\romannumeral5}, Monte Carlo simulation results are presented to show the tightness of our proposed SE lower bound. Finally, section \uppercase\expandafter{\romannumeral6} concludes this paper.

\emph{Notations}: The lowercase and uppercase boldface letters denote column vectors and matrices respectively. The operators $(\cdot)^T$, $(\cdot)^H$, $| \cdot |$ and $\|(\cdot)\|$ indicate the transposition, conjugate transposition, absolute function and Frobenius norm respectively. $\mathbf{A}(i,j)$ denotes the element of matrix $\mathbf{A}$ in $i$-th row and $j$-th column. The abbreviation $\mathbf{0}_{n\times m}$ denotes an $n$-by-$m$ zero matrix, $\mathbf{I}_{n}$ indicates an $n$-by-$n$ identical matrix, and $\mathbf{e}_n$ denotes the $n$-th column of identity matrix. The abbreviation diag$(\mathbf{x})$ denotes a diagonal matrix of which diagonal elements are $\mathbf{x}$, and $\text{det}(\mathbf{A})$ denote taking determinant of matrix $\mathbf{A}$. $\{0,1\}^N$ denotes an integer vector which is composed of $N$ elements selected from 0 and 1. $\mathcal{CN}(\boldsymbol{\mu}, \mathbf{\Sigma})$ implies a circularly symmetric multi-variate complex Gaussian distribution whose mean is $\boldsymbol{\mu}$ and covariance is $\mathbf{\Sigma}$, and $\mathcal{CN}(\mathbf{x}; \boldsymbol{\mu}, \mathbf{\Sigma})$ indicates the probability density function (PDF) of the random vector $\mathbf{x} \sim \mathcal{CN}(\boldsymbol{\mu}, \mathbf{\Sigma})$.

\section{System Model}

\subsection{Model of Uplink Single-Cell Massive SC-SM MIMO System}
\begin{figure}
  \centering
  \includegraphics[width=0.4\textwidth]{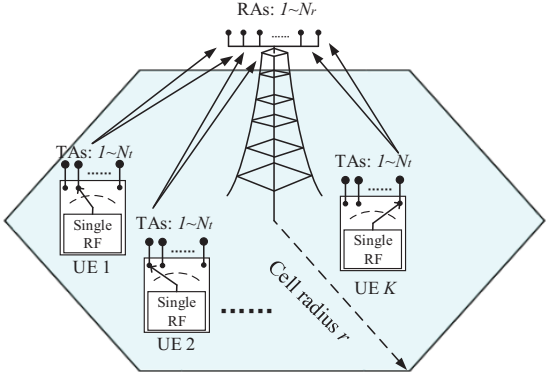}
  \caption{System model of uplink single-cell massive SC-SM MIMO system, in which BS simultaneously severs $K$ UEs under broadband transmission in the single cell of radius $r$. There are $N_r$ RAs in BS, and $N_t$ TAs of each UE.}\label{SystemModel}
\end{figure}

In this paper, as shown in Fig. \ref{SystemModel}, an uplink single-cell massive SC-SM MIMO system is introduced, in which $K$ UEs are scheduled simultaneously by the BS in the center of single cell of which radius is $r$. BS is equipped with massive RAs, and the number of RAs is $N_r$. Each UE is equipped with single RF and same number of TAs $N_t$. Since the scenario of system is broadband transmission, the dispersive channel should be regarded as a frequency-selective fading channel, rather than a Rayleigh flat fading channel. We denote $L$ as the length of multipath in this frequency-selective fading channel.

The uplink transmission is divided into frames in this system, and $N_a$ is denoted as the number of symbols in each frame, which is equal or shorter to the channel's coherence time. Therefore, within each frame, the CSI remains constant. In each frame, the former $N_c$ symbols are utilized as pilots, and next $L-1$ symbols are utilized as CP head, which are same as the last $L-1$ symbols, to eliminates the impact of CIR. So $N_s = N_a-N_c-(L-1)$ is denoted as the number of efficient symbols transmitted in each frame. With CP-aided SC-SM system, after removing the pilots and CP at the receiver, in each frame the received symbols can be formulated as follows:
\begin{equation}
\mathbf{y} = \sum_{k=1}^K\mathbf{H}_{k}\mathbf{x}_k\sqrt{P_{k}}+ \mathbf{n},
\end{equation}
where $\mathbf{y}=[\mathbf{y}_1^T,\mathbf{y}_2^T,\ldots,\mathbf{y}_{N_s}^T]^T \in\mathbb{C}^{N_rN_s\times 1}$ denotes the received efficient symbols in each frame, with $\mathbf{y}_{i}\in \mathbb{C}^{N_{r}\times 1}$ denotes the received symbols in $i$-th symbol period, $\mathbf{H}_k\in \mathbb{C}^{N_rN_s \times N_tN_s}$ is the channel matrix between $k$-th UE and BS, $\mathbf{x}_k=[\mathbf{x}_{k1}^T,\mathbf{x}_{k2}^T,\ldots,\mathbf{x}_{kN_s}^T]^T \in\mathbb{C}^{N_tN_s\times 1}$ denotes the transmit efficient symbols of $k$-th UE in each frame, with $\mathbf{x}_{ki}\in \mathbb{C}^{N_{t}\times 1}$ denotes the transmit symbols of $k$-th UE in $i$-th symbol period, $P_k$ is the transmit power of $k$-th UE to guarantee a fair service for each UE, and $\mathbf{n}\in \mathbb{C}^{N_{r}N_{s}\times 1}$ is additive white Gaussian noise (AWGN), of which all elements are independently and identically distributed (i.i.d.) $\mathcal{C}\mathcal{N}(0,\sigma_{N}^{2})$. Besides, with CP head and $L$ length multipath, the channel matrix of $k$-th UE $\mathbf{H}_{k}$ can be formulated as a block-circulant matrix, which can be formulated as follows \cite{SC-SM_PIMRC}:

\begin{equation}
\left[
    \begin{smallmatrix}
  %\begin{array}{ccccccc}
    \mathbf{H}_{k0} & \mathbf{0}  &\ldots & \mathbf{0} & \mathbf{H}_{kL-1} &\ldots & \mathbf{H}_{k1} \\
    \mathbf{H}_{k1} & \mathbf{H}_{k0} & \mathbf{0} & \ldots & \mathbf{0} & \ddots & \vdots\\
    \vdots & \ddots & \ddots & \vdots & \vdots &\ddots &\mathbf{H}_{kL-1} \\
    \mathbf{H}_{kL-1} & \mathbf{H}_{kL-2} & \ldots & \mathbf{H}_{k0} & \mathbf{0} & \ldots & \mathbf{0} \\
    \mathbf{0} & \mathbf{H}_{kL-1} & \ddots & \mathbf{H}_{k1} & \mathbf{H}_{k0} & \ddots & \vdots \\
    \vdots & \vdots & \ddots & \ddots & \ddots &\ddots &\mathbf{0} \\
    \mathbf{0} & \mathbf{0} & \ldots &\mathbf{H}_{kL-1} & \mathbf{H}_{kL-2} & \ldots & \mathbf{H}_{k0}
   \end{smallmatrix}
  %\end{array}
\right],
\end{equation}
where $\mathbf{H}_{kl}\in \mathbb{C}^{N_r\times N_t}$ denotes the channel matrix of $k$-th UE for $l$-th multipath component.

Considering the correlation between TAs, we have:
\begin{equation}
\mathbf{H}_{kl} = \mathbf{G}_{kl}\mathbf{R}_{TX}^{\frac{1}{2}},
\label{channel}
\end{equation}
where $\mathbf{R}_{TX}\in \mathbb{C}^{N_t\times N_t}$ is the transmit antenna (TA) correlation matrix. Besides, the correlation between RAs are ignored, which is because the BS has a much larger size than UEs, and the distance between RAs are much larger than that between TAs. Assuming the TAs of each UE subject to a uniform linear array, the Jakes' model \cite{Jakes} is utilized, thus we have:
\begin{equation}
\mathbf{R}_{TX}(i,j) = J_{0}(\frac{2\pi d |i-j|}{N_t\lambda}),
\end{equation}
where $J_0(\cdot)$ is the zero-order Bessel function of first kind, $d$ is the device size, and $\lambda$ is the wavelength of carrier.

As for the uncorrelated time-varying Rayleigh flat fading channel matrix $\mathbf{G}_{kl}$, of which all entries are i.i.d. complex Gaussian random variables subject to $\mathcal{C}\mathcal{N}(0,\alpha_{k}\Omega_{l})$. $\alpha_{k}$ is the large-scale attenuation between $k$th user and BS, and $\Omega_{l}$ is the power-delay profile factor as follows \cite{SC-SM_PIMRC},
\begin{equation}
\Omega_{l} = \mathbb{E}[ |\mathbf{G}_{kl}(i,j)|^{2} ] = \Omega_{0}10^{-\beta l}, l = 0,\ldots,L-1,
\end{equation}
where $\beta$ denotes the rate of decay of the average power in the multipath components in dB. And the power of multipath is normalized to unity: $\sum_{l=0}^{L-1}\Omega_{l}=1$. Thus we have the transmit power of $k$-th UE: $P_k = \frac{P_u}{\alpha_k\Omega_0}$, where $P_u > 0$ is the effective received power of each UE \cite{Massive_SM_HLZ}.

\subsection{Uplink Pilot Design and Channel Estimation}
\begin{figure}
  \centering
  \includegraphics[width=0.4\textwidth]{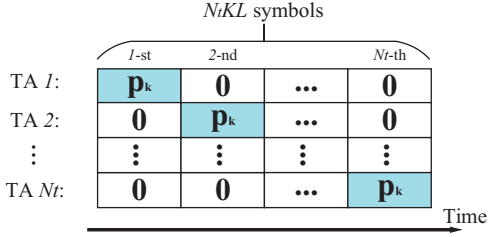}
  \caption{Orthogonal pilot design of $k$-th user, where $\mathbf{p}_k \in \mathbb{C}^{KL\times 1}$, and $\mathbf{0} \in \mathbb{C}^{KL\times 1}$ denotes the inactivting of the corresponding TA.} \label{Pilots}
\end{figure}

\begin{figure}
  \centering
  \includegraphics[width=0.4\textwidth]{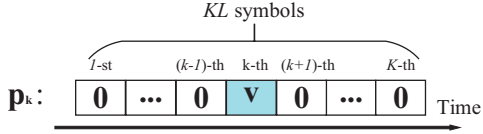}
  \caption{Pilot design for $\mathbf{p}_k$ of $k$-th user, where $\mathbf{v} \in \mathbb{C}^{L\times 1}$, and $\mathbf{0} \in \mathbb{C}^{L\times 1}$ denotes the inactivting of the corresponding TA.}\label{Pilot_p}
\end{figure}

In this paper, an orthogonal pilot design scheme and ZF channel estimation algorithm are utilized to estimate CSI. The first $N_c = L-1+N_{t}KL$ symbols are utilized as pilots, and the first $L-1$ symbols are padded as zeros to eliminate the CIR, the latter $N_{t}KL$ symbols are utilized as orthogonal pilots. As shown in Fig. \ref{Pilots} and Fig. \ref{Pilot_p}, pilot $\mathbf{p}_k$ is transmitted in $n$-th block from $n$-th TA of $k$-th user, and $\mathbf{v} = [\sqrt{KL},0,\ldots,0]^T$ is transmitted in $k$-th block of $\mathbf{p}_k$. Thus we have:
\begin{equation}
\mathbf{p}_{k}^{H}\mathbf{p}_{j} =\left\{
\begin{aligned}
& KL, & k=j, \\
& 0,  & k\neq j. \\
\end{aligned}
\right.
\end{equation}

With ZF channel estimation algorithm, the reconstructed channel vector of $k$-th UE, $0$-th multipath, and $n$-th TA can be formulated as follows,
\begin{equation}
\hat{\mathbf{h}}_{k0n} = \mathbf{h}_{k0n} + \mathbf{w}_{k0},
\end{equation}
where $\mathbf{w}_{k0}\sim\mathcal{C}\mathcal{N}(0,\frac{\sigma_{N}^{2}\alpha_{k}\Omega_{0}}{KLP_{u}}\mathbf{I}_{N_r})$ is the AWGN caused by channel estimation.

\section{Framework for SE Analysis}
To derive the SE of uplink single-cell massive SC-SM MIMO system, we first formulate the received $i$-th symbol as follows,
\begin{equation}
\mathbf{y}_{i} = \sum_{k=1}^{K}\sum_{l=0}^{L-1}\sum_{n=1}^{N_{t}}\mathbf{x}_{k,i-l,n}\sigma_{k,i-l,n}\mathbf{h}_{kln}\sqrt{P_{k}} + \mathbf{n}_i, \label{yi}
\end{equation}
where $\mathbf{n}_i \sim \mathcal{CN}(0,\sigma_{N}^{2}\mathbf{I}_{N_r})$ is AWGN, and $\sigma_{k,i-l,n}\in\{ 0,1 \}$ is a binary random variable indicating the activity of the $n$-th TA, $(i-l)$-th symbol and $k$-th user. Besides we have $\mathcal{P}(\sigma_{k,i-l,n}=1)=\frac{1}{N_t}$ and $\sum_{n=1}^{N_t}\sigma_{k,i-l,n} = 1$ because of the SM principle. With CP-aided SC-SM system, if $(i-l) \leq 0$, then we denote $(i-l) \triangleq (i-l+N_s)$.

With linear combining algorithms, we denote $\mathbf{f}_{k0n}$ as the combining vector of $k$-th UE and $n$-th TA, then the SINR of $k$-th UE, $i$-th efficient symbol and $n$-th TA, i.e. $\text{SINR}_{kin}$ can be lower bounded by (\ref{SINR}), where $E_{\mathbf{h}}\{\cdot\}$ denotes taking expectations over random relizations of channel vector $\mathbf{h}$. The denominator of $\text{SINR}_{kin}$ contains terms of ISI, equalization error and AWGN. The prove of this lower bound can refer to \cite{Massive_SINR}.
\begin{figure*}[t]
% \normalsize
\small
\begin{equation}
\text{SINR}_{kin} = \frac{\frac{P_k}{N_t}| E_{\mathbf{h}}\{ \mathbf{f}_{k0n}^H\mathbf{h}_{k0n} \} |^2}{\sum_{k'=1}^{K}\sum_{l'=0}^{L-1}\sum_{n'=1}^{N_t}\frac{P_k'}{N_t}E_{\mathbf{h}}\{ | \mathbf{f}_{k0n}^H\mathbf{h}_{k'l'n'}|^2 \} - \frac{P_k}{N_t}| E_{\mathbf{h}}\{ \mathbf{f}_{k0n}^H\mathbf{h}_{k0n} \} |^2 + \sigma_{N}^{2}E_{\mathbf{h}}\{ \| \mathbf{f}_{k0n} \|^2 \}},
\label{SINR}
\end{equation}
\begin{equation}
\text{S}_{k}^{LB} = \frac{N_s}{N_a}[ \log_2(N_t) - N_t - \frac{1}{N_t}\sum_{n = 1}^{N_t}\log_2[\sum_{m=1}^{N_t}\frac{\det(\Sigma_{kin})}{\det(\Sigma_{kin}+\Sigma_{kim})}] + \frac{1}{N_t}\sum_{n=1}^{N_t}\log_2(1+N_{t}\text{SINR}_{kin})],
\label{SE}
\end{equation}
\begin{equation}
\frac{1}{\text{SINR}_{kin}^{\text{MR}}} = \sum_{\substack{n'=1\\n'\neq n}}^{N_t}\mathbf{R}_{TX}^{2}(n',n) + \frac{N_t}{N_r}(\frac{K}{\Omega_0} - \frac{1}{N_t} + \frac{\sigma_{N}^{2}}{P_u})(1 + \frac{\sigma_{N}^{2}}{KLP_u} )
\label{SINR_MR}
\end{equation}
\hrulefill
\end{figure*}

Since we have the lower bound of $\text{SINR}_{kin}$, to derive the lower bound of mutual information $I(\mathbf{y}_{ki};\mathbf{x}_{ki})$, where $\mathbf{y}_{ki}$ is the $i$-th received symbol of $k$-th UE, we transform $\mathbf{y}_{ki}$ as the input plus additive Gaussian noise as follows,
\begin{equation}
\hat{\mathbf{y}}_{ki} = \mathbf{x}_{ki} + \mathbf{m}_{ki},
\end{equation}
where $\hat{\mathbf{y}}_{ki}\in \mathbb{C}^{N_t\times 1}$ is the equivalent $i$-th received symbol of $k$-th UE, and $\mathbf{m}_{ki}\in \mathbb{C}^{N_t\times 1}$ is a circularly symmetric complex-valued Gaussian noise with zero mean and covariance matrix as follows,
\begin{equation}
E\{\mathbf{m}_{ki}\mathbf{m}_{ki}^H\} = \text{diag}\{\frac{1}{\text{SINR}_{ki1}},\ldots,\frac{1}{\text{SINR}_{kiN_t}}\}.
\end{equation}
As the noise is modeled as additive Gaussian noise with same energy, thus we have $I(\hat{\mathbf{y}}_{ki};\mathbf{x}_{ki})\leq I(\mathbf{y}_{ki};\mathbf{x}_{ki})$.

\begin{theorem}
The lower bound SE of $k$-th UE $S_{k}^{LB}$ can be formulated as (\ref{SE}), where
\begin{equation}
\mathbf{\Sigma}_{kin} = \text{diag}\{\frac{1}{\text{SINR}_{ki1}},...,\frac{1}{\text{SINR}_{kiN_t}}\} + N_t \text{diag}\{\mathbf{e}_n\}.
\label{Sigma}
\end{equation}
\label{theorem1}
\end{theorem}

\begin{IEEEproof}
The SE $S_{k}$ of this system can be formulated as follows,
\begin{equation}
\begin{split}
S_k & = \frac{N_s}{N_a}\frac{1}{N_s}I(\mathbf{y}_{k};\mathbf{x}_{k}) \\
&\geq \frac{1}{N_a}I(\hat{\mathbf{y}}_{k};\mathbf{x}_{k}) = \frac{1}{N_a}\sum_{i = 1}^{N_s}I(\hat{\mathbf{y}}_{ki};\mathbf{x}_{ki}),
\label{Sk_sim}
\end{split}
\end{equation}
where $N_s/N_a$ is the percentage of efficient symbols, $\mathbf{y}_{k} = [\mathbf{y}_{k1}^T,\mathbf{y}_{k2}^T,\ldots,\mathbf{y}_{kN_s}^T]^T$, and $\hat{\mathbf{y}}_{k} = [\hat{\mathbf{y}}_{k1}^T,\hat{\mathbf{y}}_{k2}^T,\ldots,\hat{\mathbf{y}}_{kN_s}^T]^T$. Therefore, $S_{k}$ can be formulated as follows,
\begin{equation}
S_k \geq \frac{1}{N_a}\sum_{i = 1}^{N_s} I(\hat{\mathbf{y}}_{ki};\hat{\mathbf{e}}_{ki}) + I(\hat{\mathbf{y}}_{ki};s_{ki}|\hat{\mathbf{e}}_{ki}),
\label{S}
\end{equation}
where we have $\mathbf{x}_{ki} = s_{ki}\hat{\mathbf{e}}_{ki}$, $s_{ki}$ is the continuous complex Gaussian input, and $\hat{\mathbf{e}}_{ki} \in \{0,1\}^{N_t}$ is the active pattern of TA.

When the active TA for SM is determined, the channel can be treated as a Shannon's continuous-input continuous-output memoryless channel (CMCC), thus we have:
\begin{equation}
I(\hat{\mathbf{y}}_{ki};s_{ki}|\hat{\mathbf{e}}_{ki}) = \frac{1}{N_t}\sum_{n=1}^{N_t}\log_2(1+N_{t}\text{SINR}_{kin}).
\label{Il}
\end{equation}

Besides, as for the first term in (\ref{S}), we have:
\begin{equation}
\begin{split}
&I(\hat{\mathbf{y}}_{ki};\hat{\mathbf{e}}_{ki}) = \\ &\int\sum_{n=1}^{N_t}\mathcal{P}(\mathbf{e}_{n})\mathcal{P}(\hat{\mathbf{y}}_{ki}|\mathbf{e}_{n}) \log_2\frac{\mathcal{P}(\hat{\mathbf{y}}_{ki}|\mathbf{e}_{n})}{\sum_{m=1}^{N_t}\mathcal{P}(\mathbf{e}_{m})\mathcal{P}(\hat{\mathbf{y}}_{ki}|\mathbf{e}_{m})} d\hat{\mathbf{y}}_{ki},
\end{split}
\end{equation}
where $\mathcal{P}(\mathbf{e}_{n}) = 1/N_t$ because of the SM principle, and $\mathcal{P}(\hat{\mathbf{y}}_{ki}|\mathbf{e}_{n}) = \mathcal{CN}(\hat{\mathbf{y}}_{ki};\mathbf{0},\mathbf{\Sigma}_{kin})$. Furthermore, we have:
\begin{equation}
\begin{split}
&I(\hat{\mathbf{y}}_{ki};\hat{\mathbf{e}}_{ki}) = \\
&\frac{1}{N_t}\sum_{n=1}^{N_t}\int\mathcal{P}(\hat{\mathbf{y}}_{ki}|\mathbf{e}_{n}) \log_2\mathcal{P}(\hat{\mathbf{y}}_{ki}|\mathbf{e}_{n})d\hat{\mathbf{y}}_{ki} - \\
&\frac{1}{N_t}\sum_{n=1}^{N_t}\int\mathcal{P}(\hat{\mathbf{y}}_{ki}|\mathbf{e}_{n})\log_2 [ \frac{1}{N_t}\sum_{m=1}^{N_t}\mathcal{P}(\hat{\mathbf{y}}_{ki}|\mathbf{e}_{m})d\hat{\mathbf{y}}_{ki}] \\
& = I_{ki1} - I_{ki2}.
\end{split}
\label{Iki}
\end{equation}

As for the first term $I_{ki1}$, with $\mathcal{P}(\hat{\mathbf{y}}_{ki}|\mathbf{e}_{n}) = \mathcal{CN}(\hat{\mathbf{y}}_{ki};\mathbf{0},\mathbf{\Sigma}_{kin})$, the integration can be directly calculated out, thus we have:
\begin{equation}
I_{ki1} = -N_t\log_2(\pi e) - \frac{1}{N_t}\sum_{n=1}^{N_t}\log_2(\det(\mathbf{\Sigma}_{kin})).
\label{Iki1}
\end{equation}

To derive the closed-form solution of the second term $I_{ki2}$, Jensen's inequality is applied as follows,
\begin{equation}
I_{ki2} \leq \frac{1}{N_t}\sum_{n=1}^{N_t}\log_2[\frac{1}{N_t}\sum_{m=1}^{N_t} \int \mathcal{P}(\hat{\mathbf{y}}_{ki}|\mathbf{e}_{n}) \mathcal{P}(\hat{\mathbf{y}}_{ki}|\mathbf{e}_{m})d\hat{\mathbf{y}}_{ki}].
\label{Jensen}
\end{equation}
Then with the expression of $\mathcal{P}(\hat{\mathbf{y}}_{ki}|\mathbf{e}_{n})$, the integration can be calcuated out, thus we have:
\begin{equation}
I_{ki2} \leq -N_t\log_2\pi + \frac{1}{N_t}\sum_{n=1}^{N_t}\log_2[ \sum_{m=1}^{N_t} \frac{\frac{1}{N_t}}{\det(\mathbf{\Sigma}_{kin}+\mathbf{\Sigma}_{kim})} ].
\label{Iki2}
\end{equation}

By applying (\ref{Iki1}) and (\ref{Iki2}) into (\ref{Iki}), the lower bound of $I(\hat{\mathbf{y}}_{ki};\hat{\mathbf{e}}_{ki})$ can be formulated as follows,
\begin{equation}
\begin{split}
I(\hat{\mathbf{y}}_{ki};\hat{\mathbf{e}}_{ki}) &\geq \log_2(N_t) - N_t\log_2e\\
& -\frac{1}{N_t}\sum_{n = 1}^{N_t} \log_2[\sum_{m=1}^{N_t}\frac{\det(\mathbf{\Sigma}_{kin})} {\det(\mathbf{\Sigma}_{kin}+\mathbf{\Sigma}_{kim})}].
\end{split}
\label{I}
\end{equation}

In addition, with basic principle of SM, it can be easily derived that for each $n$, when $\text{SINR}_{kin}\rightarrow 0$, the mutual information between $\hat{\mathbf{y}}_{ki}$ and $\hat{\mathbf{e}}_{ki}$ should be $0$. When $\text{SINR}_{kin}\rightarrow \infty$, the mutual information between $\hat{\mathbf{y}}_{ki}$ and $\hat{\mathbf{e}}_{ki}$ should be $\log_2 N_t$. But with the lower bound expression of $I(\hat{\mathbf{y}}_{ki};\hat{\mathbf{e}}_{ki})$ in (\ref{I}), we have:
\begin{equation}
\begin{split}
&\lim_{\max \text{SINR}_{kin}\rightarrow 0}I(\hat{\mathbf{y}}_{ki};\hat{\mathbf{e}}_{ki}) = N_t-N_t\log_2 e \\
&\lim_{\min \text{SINR}_{kin}\rightarrow \infty}I(\hat{\mathbf{y}}_{ki};\hat{\mathbf{e}}_{ki}) =N_t-N_t\log_2 e + \log_2(N_t).
\end{split}
\end{equation}

Therefore, a constant-value is added to obtain an unbiased lower bound of $I(\hat{\mathbf{y}}_{ki};\hat{\mathbf{e}}_{ki})$ as follows,
\begin{equation}
\begin{split}
I(\hat{\mathbf{y}}_{ki};\hat{\mathbf{e}}_{ki}) &\gtrsim \log_2(N_t) - N_t \\
& -\frac{1}{N_t}\sum_{n = 1}^{N_t}\log_2[\sum_{m=1}^{N_t}\frac{\det(\mathbf{\Sigma}_{kin})}{\det(\mathbf{\Sigma}_{kin}+\mathbf{\Sigma}_{kim})}].
\end{split}
\label{Ir}
\end{equation}

With the equation (\ref{Il}) and (\ref{Ir}), the $S_k$ in (\ref{S}) can be lower bounded as (\ref{SE}), thus the proof is completed.
\end{IEEEproof}

From (\ref{SE}), it can be easily derived that a higher SINR causes a higher SE.

\section{Tight Lower Bound of SE with MR Combining}

For MR equalization, the combining vector of $k$-th UE and $n$-TA can be formulated as follows,
\begin{equation}
\mathbf{f}_{k0n}^{\text{MR}} = \hat{\mathbf{h}}_{k0n} = \mathbf{h}_{k0n} + \mathbf{w}_{k0},
\label{f}
\end{equation}
where $\mathbf{w}_{k0}\sim\mathcal{C}\mathcal{N}(0,\frac{\sigma_{N}^{2}\alpha_{k}\Omega_{0}}{KLP_{u}}\mathbf{I}_{N_r})$. With the expression of $\mathbf{f}_{k0n}^{\text{MR}}$, Theorem \ref{theorem2} can be introduced.

\begin{theorem}
The reciprocal of SINR of $k$-th UE, $i$-th symbol, and $n$-th TA in uplink single-cell massive SC-SM MIMO system with MR combining, can be formulated as (\ref{SINR_MR}).
\label{theorem2}
\end{theorem}

\begin{IEEEproof}
With the expression of (\ref{channel}) and (\ref{f}), we can easily derive out:
\begin{equation}
\begin{split}
&\mathbf{f}_{k0n}^{\text{MR}} \sim \mathcal{CN}(\mathbf{0}, \alpha_k\Omega_0(1+\frac{\sigma_{N}^{2}}{KLP_u}) \mathbf{I}_{Nr}), \\
&\mathbf{h}_{k0n} \sim \mathcal{CN}(\mathbf{0},\alpha_k\Omega_0\mathbf{I}_{Nr}).
\end{split}
\end{equation}

Thus we have:
\begin{equation}
E_{\mathbf{h}}\{ (\mathbf{f}_{k0n}^{\text{MR}})^H\mathbf{h}_{k0n} \} = \alpha_k\Omega_0N_r,
\label{E1}
\end{equation}
\begin{equation}
E_{\mathbf{h}}\{ \| \mathbf{f}_{k0n}^{\text{MR}} \|^2 \} = \alpha_k\Omega_0N_r + \frac{\sigma_{N}^{2}\alpha_k\Omega_0N_r}{KLP_u}.
\label{E2}
\end{equation}

As for the term representing ISI, only when the UE and multipath are all same, $\mathbf{f}_{k0n}^{\text{MR}}$ and $\mathbf{h}_{k'l'n'}$ are correlated because of the TA correlation. Otherwise, $\mathbf{f}_{k0n}^{\text{MR}}$ and $\mathbf{h}_{k'l'n'}$ are uncorrelated. Therefore, two cases are considered as follows:

\begin{enumerate}
\item $k'\neq k$ or $l'\neq 0$:

In this case, $\mathbf{f}_{k0n}^{\text{MR}}$ and $\mathbf{h}_{k'l'n'}$ are uncorrelated. Thus we have:
\begin{equation}
\begin{split}
&E_{\mathbf{h}}\{ | (\mathbf{f}_{k0n}^{\text{MR}})^H\mathbf{h}_{k'l'n'}|^2 \} = \alpha_{k'}\Omega_{l'}E_{\mathbf{h}}\{ \| \mathbf{f}_{k0n}^{\text{MR}} \|^2 \} \\
&=\alpha_{k'}\Omega_{l'}\alpha_k\Omega_0(N_r + \frac{\sigma_{N}^{2}N_r}{KLP_u}).
\end{split}
\label{fh}
\end{equation}

\item $k'=k$ and $l'=0$:

In this case, $\mathbf{f}_{k0n}^{MR}$ and $\mathbf{h}_{k0n'}$ are correlated because of the TA correlation. With the expression of (\ref{channel}) and (\ref{f}), we have
\begin{equation}
\begin{split}
E_{\mathbf{h}}\{ \mathbf{f}_{k0n}^{\text{MR}}\mathbf{h}_{k0n'}^H \} = \alpha_k\Omega_0\mathbf{R}_{TX}(n,n')\mathbf{I}_{Nr}, \\
E_{\mathbf{h}}\{ \mathbf{h}_{k0n'}(\mathbf{f}_{k0n}^{\text{MR}})^H \} = \alpha_k\Omega_0\mathbf{R}_{TX}(n,n')\mathbf{I}_{Nr}.
\end{split}
\end{equation}

Therefore, with the property of multi-dimensional Gaussian distribution, the conditional expectation vector and conditional covariance matrix can be derived as follows,
\begin{equation}
E\{ \mathbf{f}_{k0n}^{\text{MR}}|\mathbf{h}_{k0n'} \} = \mathbf{R}_{TX}(n,n')\mathbf{h}_{k0n'},
\end{equation}
\begin{equation}
\begin{split}
&Cov\{ \mathbf{f}_{k0n}^{\text{MR}}|\mathbf{h}_{k0n'} \} = \\ &\alpha_k\Omega_0(1-\mathbf{R}^2_{TX}(n,n')+\frac{\sigma_{N}^{2}}{KLP_u})\mathbf{I}_{N_r},
\end{split}
\end{equation}
where $Cov(\cdot)$ denotes taking covariance. Then we have:
\begin{equation}
\begin{split}
&E_{\mathbf{h}}\{ | (\mathbf{f}_{k0n}^{\text{MR}})^H\mathbf{h}_{k0n'}|^2 \}   \\
&= \mathbf{R}^2_{TX}(n,n')E_{\mathbf{h}}\{\|\mathbf{h}_{k0n'}\|^4\} + \\
&\alpha_k\Omega_0(1-\mathbf{R}^2_{TX}(n,n')+\frac{\sigma_{N}^{2}}{KLP_u}) E_{\mathbf{h}}\{\|\mathbf{h}_{k0n'}\|^2\}.
\end{split}
\end{equation}

With the property of central complex-valued Wishart distribution \cite{Wishart}, we have:
\begin{equation}
E_{\mathbf{h}}\{\|\mathbf{h}_{k0n'}\|^4\} = \alpha_k^2\Omega_0^2N_r(N_r + 1).
\label{Wis}
\end{equation}

With the expression of (\ref{E2}) and (\ref{Wis}), we can easily derive out:
\begin{equation}
\begin{split}
&E_{\mathbf{h}}\{ | (\mathbf{f}_{k0n}^{\text{MR}})^H\mathbf{h}_{k0n'}|^2 \}   \\
&=\alpha_k^2\Omega_0^2N_r(N_r\mathbf{R}^2_{TX}(n,n') + 1 + \frac{\sigma_{N}^{2}}{KLP_u}).
\end{split}
\label{fh2}
\end{equation}
\end{enumerate}

With the equation (\ref{E1}), (\ref{E2}), (\ref{fh}) and (\ref{fh2}), the reciprocal of $\text{SINR}_{kin}^{\text{MR}}$ can be derived as (\ref{SINR_MR}), thus the proof is completed.
\end{IEEEproof}

With Theorem \ref{theorem2}, by applying (\ref{SINR_MR}) into (\ref{Sigma}) and (\ref{SE}), the lower bound of SE in uplink single-cell massive SC-SM MIMO system with MR combining can be derived.

From the expression in (\ref{SINR_MR}), it can be shown that for MR combining, a higher TA correlation can lower the SINR. Besides, more UEs can lower the SINR too. If the dominant path has more power, the SINR will be higher. Increasing signal-to-noise ratio (SNR) can increase the SINR, but it has a limitation because of the ISI and the TA correlations. In addition, increasing $N_r$ can also increase the SINR, but it also has a limitation as follows,
\begin{equation}
\lim_{N_r\rightarrow \infty}\frac{1}{\text{SINR}_{kin}^{\text{MR}}} = \sum_{\substack{n'=1\\n'\neq n}}^{N_t}\mathbf{R}_{TX}^{2}(n',n)
\end{equation}

\section{Simulation Results}
In this section, both our proposed theoretical lower bound and simulation results of SE in uplink single-cell massive SC-SM MIMO system with MR combining are provided. The lower bound of SE is $\text{S}^{\text{LB}} = \sum_{k=1}^K \text{S}_{k}^{\text{LB}}$, and simulation of SE is $\text{S} = \sum_{k=1}^K \text{S}_{k}$, where $\text{S}_{k}^{\text{LB}}$ is calculated as (\ref{SE}) with MR combining, and $\text{S}_{k}$ is calculated as (\ref{Sk_sim}) with MR combining. As for the system parameters, we choose $N_a = 2048$ as the length of frame, $5$ GHZ as the frequency of carrier, $0.1$ m as the device size, $\beta = 3$ dB as the decay of multipath components, $\sigma_{N}^{2} = 0$ dB as the power of noise, $r = 500$ m as the cell radius, $r_m = 50$ m as the minimum distance between UEs and BS, and $\gamma = 3.7$ as the path loss exponent. We denote $v_k$ as the distance between $k$-th UE and BS, thus we have $\alpha_k = (v_k/r_m)^{-\gamma}$, and $r_m\leq v_k \leq r$.

\begin{figure}
  \centering
  \includegraphics[width=0.4\textwidth]{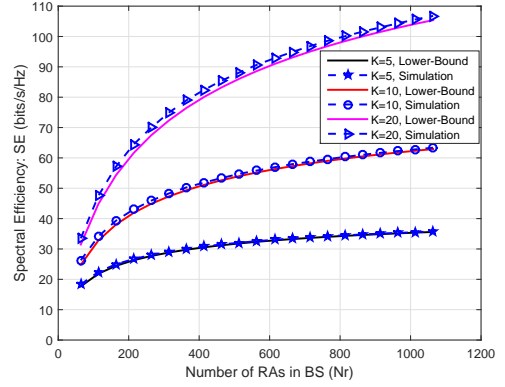}
  \caption{SE performance versus $N_r$ with $N_t = 2$, $K = 5,10,20$, $L = 3$ and $P_u = 10$ dB.}\label{Simulation_Nr}
\end{figure}

\begin{figure}
  \centering
  \includegraphics[width=0.4\textwidth]{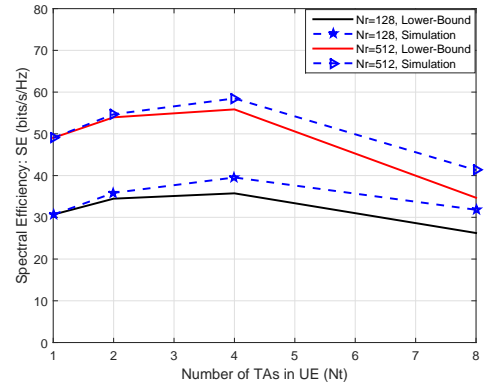}
  \caption{SE performance versus $N_t$ with $N_r = 128,512$, $K = 10$, $L = 3$ and $P_u = 10$ dB.}\label{Simulation_Nt}
\end{figure}

As shown in Fig. \ref{Simulation_Nr}, with the increase of $N_r$, a higher SE can be achieved, which is because a larger $N_r$ brings a higher receive diversity. In Fig. \ref{Simulation_Nt}, with the increase of $N_t$, the SE firstly increases then decreases. This is because a larger $N_t$ brings more spatial information, but with more ISI, TA correlations and pilots. It is worth noting that in this circumstance, the uplink single-cell massive SC-SM MIMO system outperforms the single TA counterpart concerning SE.

\begin{figure}
  \centering
  \includegraphics[width=0.4\textwidth]{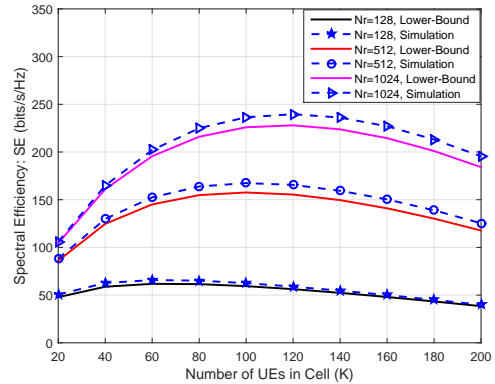}
  \caption{SE performance versus $K$ with $N_t = 2$, $N_r = 512$, $L = 3$ and $P_u = 10$ dB.}\label{Simulation_K}
\end{figure}

\begin{figure}
  \centering
  \includegraphics[width=0.4\textwidth]{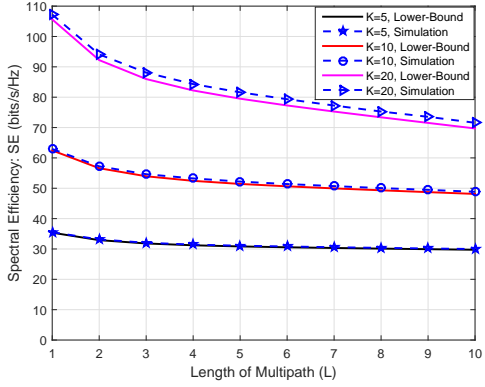}
  \caption{SE performance versus $L$ with $N_t = 2$, $N_r = 512$, $K = 5,10,20$ and $P_u = 10$ dB.}\label{Simulation_L}
\end{figure}

\begin{figure}
  \centering
  \includegraphics[width=0.4\textwidth]{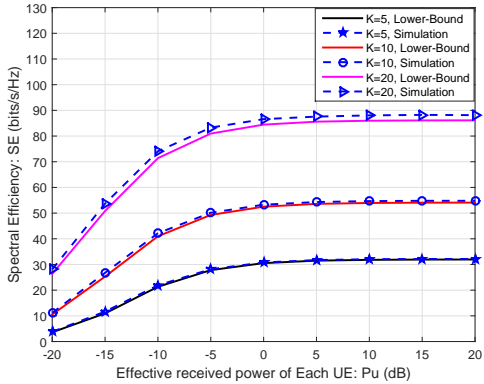}
  \caption{SE performance versus $P_u$ with $N_t = 2$, $N_r = 512$, $K = 5,10,20$ and $L = 3$.}\label{Simulation_Pu}
\end{figure}

In Fig. \ref{Simulation_K}, the natural tradeoff between the number of UEs and the channel estimation overhead is more intuitively shown. On the one hand, increasing $K$ leads to increasing the uplink throughput. On the other hand, when $K$ is continuously enlarged, the SE gain can no longer compensate for the drastically increasing pilot overhead and ISI, which nevertheless leads to reducing the uplink SE. In addition, the optimal $K$ maximizing the uplink SE is dependent on various system parameters. A larger $N_r$ causes a larger optimal $K$. From Fig. \ref{Simulation_L}, we can see that a larger $L$ causes a lower SE, which is because with the increase of $L$, the length of pilots becomes longer, and ISI becomes higher. As shown in Fig. \ref{Simulation_Pu}, when $P_u$ is small, it is a power-limited system, so increasing $P_u$ can increase SE. When $P_u$ is larger than $0$ dB, it becomes an interference-limited system, that is, the TA correlation and ISI mostly contributes to affect the SE, so SE almost keeps unchanged with the increasing of $P_u$.

In addition, from Fig. \ref{Simulation_Nr}, Fig. \ref{Simulation_L} and Fig. \ref{Simulation_Pu}, it can be shown that the lower bound of SE is very tight. As for the circumstances in Fig. \ref{Simulation_K}, the SE lower bound
of each UE is still tight, but the sum operation of $K$ UEs magnifies the error. In Fig. \ref{Simulation_Nt}, with the increase of $N_t$, gap between the SE lower bound and real SE becomes larger.

\section{Conclusions}
In this paper, a framework for SE analysis in uplink single-cell massive SC-SM MIMO system with linear combining algorithms is proposed, and a novel closed-form lower bound of SE with MR combining is derived. The tightness of our proposed SE lower bound is validated via Monte Carlo simulations, and we analyze the impact of system parameters on SE. Besides, both the theoretical analysis and simulation results show that the SE of uplink single-cell massive SC-SM MIMO system has the potential to outperform the SE of single antenna counterpart.


\begin{thebibliography}{1}

\bibitem{Massive_CM}
E. G. Larsson, O. Edfors, and F. Tufvesson, ``Massive MIMO for next generation wireless systems,'' \emph{IEEE Commun. Mag.}, vol. 52, no. 2, pp. 186-195, Feb. 2014.

\bibitem{Massive_SPM}
F. Rusek, D. Persson, B. K. Lau, E. G. Larsson, T. L. Marzetta, O. Edfors, and F. Tufvesson, ``Scaling up MIMO: Opportunities and challenges with very large arrays,'' \emph{IEEE Signal Process. Mag.}, vol. 30, no. 1, pp. 40-60, Jan. 2013.

\bibitem{Massive_SINR}
E. Bj\"{o}rnson, E. G. Larsson, and M. Debbah, ``Massive MIMO for maximal spectral efficiency: how many users and pilots should be allocated?'' \emph{IEEE Trans. Wireless Commun.}, vol. 15, no. 2, pp. 1293-1308, Feb. 2016.

\bibitem{Massive_Ding}
L. Liang, W. Xu, and X. Dong, ``Low-complexity hybrid precoding in massive multiuser MIMO systems,'' \emph{IEEE Wireless Commun. Lett.}, vol. 3, no. 6, pp. 653-656, Dec. 2014.

\bibitem{SM}
R. Mesleh, H. Haas, S. Sinanovic, C. W. Ahn, and S. Yun, ``Spatial modulation,'' \emph{IEEE Trans. Veh. Technol.}, vol. 57, no. 4, pp. 2228-2241, Jul. 2008.

\bibitem{SM_mgz}
M. Di Renzo, H. Haas, A. Ghrayeb, S. Sugiura, and L. Hanzo, ``Spatial modulation for generalized MIMO: challenges, opportunities and implementation,'' \emph{Proceedings of the IEEE}, vol. 102, no. 1, pp. 56-103, Jan. 2014.

\bibitem{Massive_SM_ITA} %HeHe
T. Narasimhan, P. Raviteja, and A. Chockalingam, ``Large-scale multiuser SM-MIMO versus massive MIMO,'' in \emph{Proc. ITA}, pp. 1-9, Feb. 2014.

\bibitem{Massive_SM_TWC}
S. Wang, Y. Li, M. Zhao, and J. Wang, ``Multiuser detection in massive spatial modulation MIMO with low-resolution ADCs,'' \emph{IEEE Trans. Wireless Commun.}, vol. 14, no. 4, pp. 2156-2168, Apr. 2015.

\bibitem{Massive_SM_HLZ}
L. He, J. Wang, and J. Song, ``On massive spatial modulation MIMO: spectral efficiency analysis and optimal system design,'' in \emph{Proc. IEEE GLOBECOM}, 2016, accepted.

\bibitem{SC-SM_Hanzo}
S. Sugiura and L. Hanzo, ``Single-RF spatial modulation requires singlecarrier transmission: frequency-domain turbo equalization for dispersive channels,'' \emph{IEEE Trans. Veh. Technol.}, vol. 64, no. 10, pp. 4870-4875, Oct. 2015.

\bibitem{SC-SM_PIMRC}
P. Som and A. Chockalingam, ``Spatial modulation and space shift keying in single carrier communication,'' in \emph{IEEE 23rd International Symposium on Personal Indoor and Mobile Radio Communications (PIMRC)}, pp. 1962-1967, Sep. 2012.

\bibitem{SC-SM_ZP}
R. Rajashekar, K. V. S. Hari, and L. Hanzo, ``Spatial modulation aided zero-padded single carrier transmission for dispersive channels,'' \emph{IEEE Trans. Commun.}, vol. 61, no. 6, pp. 2318-2329, Jun. 2013.

\bibitem{SC-SM_LS_Mag}
P. Yang, \textit{et al.}, ``Single-carrier spatial modulation: a promising design for large-scale broadband antenna systems,'' \emph{IEEE Commun. Surveys Tutorials}, vol. 18, no. 3, pp. 1687-1716, Feb. 2016.

\bibitem{SC-GSM_LS}
T. L. Narasimhan, P. Raviteja, and A. Chockalingam, ``Generalized spatial modulation in large-scale multiuser MIMO systems,'' \emph{IEEE Trans. Wireless Commun.}, vol. 24, no. 7, pp. 3764-3779, Jul. 2015.

\bibitem{SC-SM_LS_DT}
S. Wang, Y. Li, M. Zhao, and J. Wang, ``Energy efficient and lowcomplexity uplink transceiver for massive spatial modulation MIMO,'' \emph{IEEE Trans. Veh. Technol.}, vol. 64, no. 10, pp. 4617-4632, Oct. 2015.

\bibitem{Jakes}
P.-H. Kuo, H. T. Kung, and P.-A. Ting, ``Compressive sensing based channel feedback protocols for spatially-correlated massive antenna arrays,'' in \emph{Proc. IEEE Wireless Commun. Netw. Conf.}, pp. 492-497, Apr. 2012.

\bibitem{Wishart}
A. M. Tulino and S. Verd\'{u}, ``Random matrix theory and wireless communications,'' \emph{Foundations Trends Commun. Inf. Theory}, vol. 1, no. 1, pp. 1-182, Jun. 2004.

\end{thebibliography}
\end{document}